\documentclass[12pt,preprint]{aastex}

\begin{document}

\title{The Relation between Near-Infrared Luminosity of RGB Bumps
and Metallicity of Galactic Globular Clusters}

\author{Dong-Hwan Cho and Sang-Gak Lee}
\affil{Astronomy Program, School of Earth and Environmental
Sciences, \\Seoul National University, Seoul 151-742, Korea}
\email{chodh@astro.snu.ac.kr, sanggak@astrosp.snu.ac.kr}

\begin{abstract}
Using photometric data from the 2MASS second incremental release
point source catalog we
constructed $K_{s}$ vs. ($J-K_{s}$) color-magnitude diagrams (CMDs)
of Galactic globular clusters (GGCs) for which the $JK_{s}$
photometric data have been made available up to now. On the CMDs of 13 GGCs
we identified RGB bump features and derived luminosities of
bumps in $K_{s}$ ($K_{s}^{\rm Bump}$) and in $M_{K_{s}}$
($M_{K_{s}}^{\rm Bump}$) for 11 of them. We reconfirm the
relation between $M_{K_{s}}^{\rm Bump}$ and metallicity
[Fe/H]$_{\rm CG97}$ or  [M/H] such that the luminosity of bump
becomes brighter as metallicity [Fe/H]$_{\rm CG97}$ or [M/H]
decreases. This result is similar to that obtained by Ferraro et
al. (1999) for the relation between $M_{V}^{\rm
Bump}$ and [M/H] based on observations of 47 GGCs which were conducted
in the optical region (in $V$ vs.
($B-V$) CMDs). Our results show the same trend as those found by
Ferraro et al. (2000) for the
relations between $M_{K}^{\rm Bump}$ and metallicity [Fe/H]$_{\rm
cg97}$ and [M/H] derived from observations of 8 GGCs in
the same near-infrared region ($K$ vs. ($J-K$) CMDs). Combining
with the data of Ferraro et al. (2000), we
derive a robust relation for the metal-dependent luminosities of
the bumps.

\end{abstract}

\keywords{globular clusters: general---stars: color-magnitude
diagrams---stars: luminosity function---stars: evolution---stars:
population II}

\section{Introduction}

The red giant branch (RGB) bump in  globular clusters was
theoretically predicted by Thomas (1967) and Iben (1968) as a
region where evolution through the RGB is stalled for a time when
the H burning shell passes the H abundance inhomogeneity envelope.
This is produced by the stellar outer convection zone at the H
shell burning RGB stage after the first dredge-up of a star in
a globular cluster. The first convincing identification of the
bump was that of the metal-rich cluster 47 Tuc (King, Da Costa, \&
Demarque 1985).
The position in luminosity of the RGB bump is a function of metal
abundance, helium abundance, and stellar mass (and hence cluster
age) as well as any additional parameters that determine the
maximum inward extent of the convection envelope or the position
of the H burning shell. Thus the position of this peak on the
giant branch of GGCs should provide observational constraints on
these parameters.

RGB bumps in Galactic globular clusters (GGCs) were first
observationally identified in a systematic way  by Fusi
Pecci et al. (1990). They found RGB bumps in 11 GGCs from the
peaks in the differential luminosity function and a change in the
slope of the integrated luminosity function using $V$ vs. ($B-V$)
color-magnitude diagrams (CMDs). They discovered a correlation
between the positions of RGB bumps in
the GGCs and their metallicities. Subsequent to this work, many studies
on this subject have been carried out (Sarajedini \& Norris 1994;
Brocato et al. 1996; Saviane et al. 1998). Recently, Ferraro et
al. (1999) systematically identified the positions of RGB bumps in 47
GGCs in $V$ vs. ($B-V$) CMDs and confirmed the strong correlation
between the absolute $V$ magnitudes of RGB bumps and metallicities of
GGCs.

In the near-infrared band, the the locations of RGB bumps were identified
for 8 GGCs in $K$ vs. ($J-K$) CMDs and a similar correlation
between absolute $K$ magnitudes of RGB bumps and metallicities
was derived by Ferraro et al. (2000).

The advantage of observing GGCs in the near-IR is an enormous
reduction in reddening. The interstellar reddening in the $K$ band
($A_{K}$) is roughly 10$\%$ of that in the visual ($A_{V}$) part of 
the spectrum.
This imparts a great advantage to near-infrared
observations especially toward the disk and severely
obscured bulge directions. Therefore, near-infrared observations are
much preferred for disk and severely obscured bulge GGCs and
for as yet unknown GGCs severely reddened by the Galactic bulge in
the visual bands. Near-infrared observations are also useful for
normal disk and halo GGCs because the $(V-K)$ color index is a good
temperature indicator.

In the near-infrared band, RGB bumps of GGCs can be a very useful
distance indicator because they are relatively bright, and their
position in  CMDs is close to the HB (Horizontal Branch).
Moreover, the HB, which is a good distance indicator in
the $V$ band is not useful in the near-infrared CMDs. In the near-IR
CMDs, the HB is diagonally slanted from the upper right to the
lower left position and is not horizontal at all.
So it is difficult to measure the HB levels of GGCs in the near-IR CMDs.
It is expected that the RGB bump
is a distance indicator which would have the same amount of observational error
as the HB used in the optical region since it has a similar HB metallicity
dependency.

The goal of this work is to obtain the absolute magnitudes of
RGB bumps in as many GGCs as possible in the near-IR band and to
derive a robust relation between the luminosities of the bumps and
metallicities of GGCs for use as distance indicators in the IR
bands. Using the photometric data from the 2MASS (Two Micron All Sky
Survey) second incremental release point source catalog we have
investigated the existence of RGB
bumps in GGCs and have found a correlation between the
absolute magnitudes of RGB bumps and metallicities of GGCs,
comparing our findings with the results of Ferraro et al. (2000).
Combining with their data we provide a more robust
correlation for the use of RGB bumps as standard candles for
deriving distances to other GGCs.

In $\S$ 2 we discuss some characteristics of GGCs' near-infrared
CMDs and in $\S$ 3 investigate the location in luminosity of the RGB bumps and
the relation between the luminosity of the bump and metallicity
of GGCs, and in $\S$ 4 we briefly summarize our results.

\section{IR Color-Magnitude Diagrams}

We used photometric data from the 2MASS second incremental release
point source catalog which were obtained at two 1.3-m
telescopes in the northern and southern hemispheres in the
near-infrared bands at $J$ (1.25$\mu$m), $H$ (1.65$\mu$m), and 
$K_{s}$ (2.17$\mu$m; $K$ short = medium modified $K$). Limiting
magnitudes of the 2MASS second incremental release point source
catalog photometric data are about 15 mag and the released data
from 2MASS now covers $\sim$47$\%$ of the sky.

The $JK_{s}$ photometric data for GGCs in Harris' (1996) catalog
which are relatively close to the Sun are obtained from the
IPAC (Infrared Processing and Analysis Center) and $K_{s}$ vs.
($J-K_{s}$) CMDs are constructed. We rejected data whose
errors  are larger than 0.15 mag. We can construct CMDs of about
46.8$\%$ of GGCs in Harris' (1996) catalog from the 2MASS second
incremental release point source catalog. However, the  quality of many
CMDs is not good enough to find the bumps because the limiting magnitudes
are too bright.

In Figure 1 we present
CMDs of 11 GGCs for which the RGB bump positions are accurately
measured by the systematic analytical method described in $\S$ 3.1.
According to Figure 1 the major characteristics of the CMDs of GGCs
in the $K_{s}$ vs. $(J-K_{s})$ plane are as follows.

\placefigure{fig1}

First, the BHB (Blue Horizontal Branch) and RHB (Red Horizontal Branch)
are diagonally tilted from the upper right to the lower left position
and such features are obvious in the CMDs of M4 (NGC 6121)
and M107 (NGC 6171) which have both BHB
and RHB.
Even the RHB of 47 Tuc (NGC 104), which has only RHB,
is clearly tilted. So in the $K_{s}$ vs. $(J-K_{s})$
CMDs it is difficult to determine the HB level of any GGC and
to correlate RGB bump positions and HB levels in the near-infrared CMDs,
in contrast to the case for optical CMDs.

Second, the brightness interval between the MSTO (Main Sequence
Turnoff) and RGB tip is larger ($\sim$9 mag) than in optical CMDs
($\sim$6.5 mag), as can be easily seen in M4.
So the magnitude resolution in the $K_{s}$ vs. $(J-K_{s})$ CMDs
is larger than in optical CMDs.
However, color resolution in the $K_{s}$ vs. $(J-K_{s})$
CMDs is smaller than in optical CMDs because in the latter the RGB
and AGB (Asymptotic Giant Branch) are separated from one another
at least in the lower part of the AGB. In contrast, in the
$K_{s}$ vs. $(J-K_{s})$
CMDs the RGB and AGB overlap, as clearly shown in the CMD of
47 Tuc whose AGB is relatively rich. In the optical
CMDs of NGC 362, 47 Tuc, and M71 (NGC 6838), the separation of
their RGBs and RHBs is clearly seen but they partially overlap in
the $K_{s}$ vs. $(J-K_{s})$ CMDs. This also results from the fact
that the color resolution in the $K_{s}$ vs. $(J-K_{s})$ CMDs is lower
than that in optical CMDs.

Third, in the case of M22 (NGC 6656), its RGB is broader than those
of other GGCs as found in the optical CMD (Peterson \& Cudworth 1994).
Moreover, in the lower part of the RGB, contamination from bulge
component stars is severe. To the right of the RGB of M22
there exists another RGB component. This RGB component would be
due to bulge RGB stars, which extend prominently redward in the optical
CMD (in our unpublished data) because of the strong blanketing effects of
heavy metals as found in metal-rich GGCs such as
NGC 6553 (Ortolani, Barbuy, \& Bica 1990).

Last, because NGC 362 and 47 Tuc CMDs are located in the direction
of the SMC, a contribution from the SMC appears in the lower right
region of their CMDs.

\section{The RGB Bump}

\subsection{The Luminosity of the RGB Bump}

In order to accurately measure the luminosity of the RGB bump and
construct luminosity functions of the GGCs, we applied several
standard procedures to delineate only the RGB sequences for all GGCs
except for M22 and M71. First we rejected visually clear HB stars
and AGB stars in the process of distinguishing outlying field stars
from RGB stars. Second, by binning the RGB sequences in 0.5 mag
intervals we measured the average and sigma of each bin and
rejected stars 2$\sigma$ away from the mean value of each bin. We
then remeasured the average and sigma of each bin and further rejected
stars 2$\sigma$ away from the remeasured average of each bin, and
iterated this process until no stars were rejected and the average
and sigma of each bin  converged. Third, from each RGB
sequence after field star rejection, we constructed a differential
luminosity function and an integral luminosity function for each GGC
following the classical method of Fusi Pecci et al. (1990).

Then, we searched for a significant peak in the differential
luminosity function
and a corresponding large slope change in the integral luminosity function
simultaneously for each GGC. Where the significant peak position in the
differential luminosity function and the position of large slope change
in the integral luminosity function coincide, we identified the
significant peak as the RGB bump of the given GGC and measured this
position ($K_{s}^{\rm Bump}$). However, in several GGCs the existence of
RGB bumps is very clear only when considering the differential luminosity
functions.

In the case of M22, since its CMD is severely contaminated by field and
bulge stars, we clipped off the proper region
in order to isolate the RGB sequence to $K_{s}$ $=$ 13.
Below $K_{s}$ $=$ 13, isolation of the RGB sequence is difficult
because of severe contamination by the field and bulge stars.

In the case of M71, the RGB is too poorly populated to be clearly
defined. So in order to isolate the RGB sequence of
M71 we used the 2$\sigma$ clipped RGB sequence of 47 Tuc with a similar
metallicity as a template for the RGB sequence of M71. The
remaining processes were the same as for the other GGCs.

Differential luminosity functions and integral luminosity functions
for each GGC are shown in Figure 2.

\placefigure{fig2}

In order to derive absolute magnitudes of GGC RGB bumps
($M_{K_{s}}^{\rm Bump}$) from their apparent magnitudes
we need absolute distance moduli and interstellar reddenings.
We take these values from Table 2 of Ferraro et al.
(1999) except for M22 and M2 (NGC 7089). Ferraro et al. (1999)
derived absolute distance moduli in a very systematic way using
the HB level as a distance indicator, or more strictly speaking the ZAHB (Zero
Age Horizontal Branch) level. They derived ZAHB levels for GGCs by
comparing the CMDs of GGCs with synthetic H-R diagrams. Then they
derived absolute distance moduli using absolute magnitude levels
of the ZAHB ($M_{V}^{\rm ZAHB}$) and interstellar reddening. For
interstellar reddening they adopted values from the compilation of
Harris (1996).

In the cases of M22 and M2, which are not in the list of Ferraro
et al. (1999), we calculated their absolute distance moduli based
on the same method described in Ferraro et al. (1999), referring
to other sources for the physical values. For derivation of the absolute
distance moduli of M22 and M2 we used the magnitude difference between
the apparent
visual magnitude of the ZAHB ($V_{\rm ZAHB}$) and the absolute visual magnitude
of the ZAHB ($M_{V}^{\rm ZAHB}$) of each cluster. That is as follows:

\begin{equation}
(m-M)_{0}^{\rm CG97} = (m-M)_{V} - A_{V} =
(V_{\rm ZAHB} - M_{V}^{\rm ZAHB}) - A_{V} .
\end{equation}

For $V_{\rm ZAHB}$ we used equation (2) in $\S$ 5 of Ferraro et al.
(1999):

\begin{equation}
V_{\rm ZAHB} = {<}V_{\rm HB}{>} + 0.106{\rm [M/H]}^{2} +
0.236{\rm [M/H]} + 0.193 .
\end{equation}

By the way, $\alpha$-element enhanced global metallicity [M/H]
values for M22 and M2
were computed by equation (1) in $\S$ 3.4 of Ferraro et al. (1999):

\begin{equation}
{\rm [M/H]} = {\rm [Fe/H]} + {\rm log} (0.638f_{\alpha} + 0.362) ,
\end{equation}

where $f_{\alpha}$ is the enhancement factor of the $\alpha$-elements.
According to Ferraro et al. (1999), for the 19 GGCs with [$\alpha$/Fe]
listed by Salaris \& Cassisi (1996) or Carney (1996), they adopted
$f_{\alpha}$ = $10^{\rm [{\alpha}/Fe]}$.
For all the others they assumed
$f_{\alpha}$ = $10^{0.28}$ if [Fe/H] $<$ $-0.8$ and $f_{\alpha}$ =
$10^{\rm -0.35[Fe/H]}$ if [Fe/H] $>$ $-0.8$. So, for M22 we adopted
[$\alpha$/Fe] = 0.32 from Carney (1996), averaged over 9 stars which include
all 3 stars of M22 in Salaris \& Cassisi (1996). For M2 [$\alpha$/Fe] is
not listed in Salaris \& Cassisi (1996) or Carney (1996). So, we adopted
$f_{\alpha}$ = $10^{0.28}$ because [Fe/H] of M2 is $-1.46$. For
$M_{V}^{\rm ZAHB}$ we used equation (4) for [Fe/H] in $\S$ 6.4 of
Ferraro et al. (1999):

\begin{equation}
M_{V}^{\rm ZAHB} = 1.0005 + 0.3485{\rm [Fe/H]} +
0.0458{\rm [Fe/H]}^{2} .
\end{equation}

For $A_{V}$ we adopted $A_{V}$ = 3.08$E(B-V)$ from He et al. (1995)
and for $A_{K_{s}}$, which is used in the calculation of apparent
$K_{s}$ magnitude distance moduli for program GGCs, we adopted
$A_{K_{s}}$ = 0.32$E(B-V)$ from He et al. (1995). All parameters
of M22 and M2 are listed in Table 1.

\placetable{tbl-1}

All these final values for absolute distance moduli and
interstellar reddening of program GGCs are listed in Table 2. According to
Ferraro et al. (1999) the global uncertainty of the absolute distance
moduli is of the order of about 0.2 mag considering that the derived
absolute distance moduli are affected by many uncertainties (namely,
the evaluation of the ZAHB level, the zero point and
dependence on metallicity of the ZAHB level, reddening, etc.).
Finally, using absolute distance moduli and interstellar reddenings
we derived absolute magnitudes of the RGB bumps of GGCs
($M_{K_{s}}^{{\rm Bump}}$) listing them in Table 2 together with
$K_{s}^{\rm Bump}$. In Table 2, errors in column (6) are
measurement errors and errors in column (7) are a combination
of measurement errors in column (6) and the global uncertainty of the
absolute distance moduli, which is  equal to 0.2 mag.

\placetable{tbl-2}

In the case of M2, the RGB bump appears in the expected position
according to equations (5a), (5b) and (6a), (6b) in the next section,
but the bump is relatively broader compared to those of other GGCs, and the
quality of its IR CMD is poor compared to the optical CMD of
Lee \& Carney (1999).

Cudworth (1985) claimed that in the optical CMD of M71 there is a strong
clump just below the HB whose position in the $V$ vs. $(B-V)$
plane is $V$ $\approx$ 15 and $(B-V)$ $\approx$ 1.2. Transforming
these values into the $J$ vs. $(J-K_{s})$ plane by the color
transformation table of Bessell, Castelli, \& Plez (1998), $V$ $\approx$ 15
transforms into $K_{s}$ $\approx$ 11.92 and $(B-V)$ $\approx$ 1.2
transforms into $(J-K_{s})$ $\approx$ 0.77. Also, according to
Girardi et al.'s (2000) evolutionary tracks
the former values correspond to $K_{s}$ $\approx$ 11.94
and $(J-K_{s})$ $\approx$ 0.72. These transformed $K_{s}$ and $(J-K_{s})$
values of the strong clump of M71 in the $V$ vs. $(B-V)$ CMD
nearly match the RGB bump position
$K_{s}$ $=$ 11.95 and $(J-K_{s})$ $=$ 0.72 derived in this work. So
the strong clump reported by Cudworth (1985) in the $V$ vs. $(B-V)$ CMD
of M71 must be the RGB bump derived in this work
in the $K_{s}$ vs. $(J-K_{s})$ CMD.

In the cases of M69 (NGC 6637) and $\omega$ Cen (NGC 5139) we can see
the RGB bump features in their CMDs. However, for M69 it is
difficult to measure the RGB bump position by systematic analysis
because it exhibits a CMD which is too scatted to analyze.
Also for $\omega$ Cen we abandoned the analysis because
$\omega$ Cen has a
wide RGB suggesting a metallicity spread
(Lee et al. 1999; Pancino et al. 2000), which renders measurement of
RGB bump position meaningless in this work.

However, in the cases of M30 (NGC 7099) and M55 (NGC 6809), we cannot
detect RGB bumps at the expected positions taking into account their
metallicity after careful analysis of their differential
and integral luminosity functions. Figure 3 shows their $K_{s}$ vs.
$(J-K_{s})$ CMDs, differential luminosity functions, and integral
luminosity functions.
For M30 there are hints of bumps at $K_{s}$ $\sim$ 12.5 and
$\sim$ 13.5. However, no slope changes around those magnitudes are found
in the integral luminosity function. For M55 neither a bump nor
a significant slope change are found.

\placefigure{fig3}

\subsection{The Relation between the luminosity of RGB bump and metallicity}

There are two widely used metallicity scales, the Zinn \& West scale
(Zinn \& West 1984; Zinn 1980, 1985; Armandroff \& Zinn
1988) and the Carretta \& Gratton scale (Carretta \& Gratton 1997).
The Zinn \& West scale (hereafter ZW scale) employs the most complete data
set but is not based on high resolution spectra. However, the
Carretta \& Gratton scale (hereafter CG97 scale) is based on
systematic high resolution spectra obtained by their own team, and
there is a transformation relation between the ZW scale and the CG97
scale in equation (7) of Carretta \& Gratton (1997). The CG97
scale is more robust since it relies on recent high resolution
spectroscopic measurements and up-to-date atmospheric models.
Moreover, in order to compare our results directly with those of Ferraro et
al. (2000) which employed the CG97 scale, we adopted the CG97
metallicity scale.

The CG97 scale metallicity of each GGC is given in Table 2. Since
only 24 GGCs have metallicity determinations from direct
high resolution spectra in the CG97 scale, the metallicities of NGC 1851
and M107 are interpolated from equation (7) of Carretta \&
Gratton (1997), and the metallicity of M2 is adopted from the value
determined by the
morphological parameters of RGB stars by Lee \& Carney (1999).
Recently, Carretta \& Gratton's group published high resolution spectroscopic
metallicities of NGC 6553 (Cohen et al. 1999; Carretta et al 2001) and
NGC 6528 (Carretta et al. 2001, hereafter CG), extending the CG97 metallicity
scale into the
high metallicity region ([Fe/H]$_{\rm CG}$ $\approx$ 0.00), and established
a new transformation equation between ZW scale metallicity and CG scale
metallicity. However, in the metallicity range of [Fe/H]$_{\rm CG97}$ $= -2.12
\sim -0.70$ the interpolated metallicity values are coincident within 0.1 dex
in the two different transformation equations and it seems that transformation
equation (7) of Carretta \& Gratton (1997) is tighter than transformation
equation (3) of Carretta et al. (2001). We did not revise the metallicities
of NGC 1851, M107, M2, M55, and M69, which have no metallicity measurements
from direct high resolution spectra in the CG97 metallicity scale. But for
metallicities of NGC 6553 and NGC 6528, which were extrapolated by
Ferraro et al. (1999) since their ZW scale metallicities  are outside
the validity range of the transformation to the CG97 scale metallicities by
equation (7) of Carretta \& Gratton (1997), we adopted the new high resolution
spectroscopic measurements of Carretta et al. (2001).
Metallicities of the other GGCs are taken from the directly
determined values of Carretta \& Gratton (1997).

We also consider
the global metallicity [M/H] which incorporates $\alpha$-element
enhancement into the CG97 metallicity scale [Fe/H]$_{\rm CG97}$
according to equation (1) of Ferraro et al. (1999).
In the case
of [M/H] we directly adopted the values in Table 1 of Ferraro et al.
(1999) except for M22, M2, NGC 6553, and NGC 6528. In the cases of M22
and M2 we calculated [M/H] as described in $\S$ 3.1 and in the cases of
NGC 6553 and NGC 6528 we first calculated [$\alpha$/Fe] using equation
(2) of Carney (1996) with the $\alpha$-element abundances listed in
Carretta et al. (2001) and then calculated [M/H] using equation (3) in
$\S$ 3.1. Values of [$\alpha$/Fe] of NGC 6553 and NGC 6528 are
0.20 $\pm$ 0.07 and 0.21 $\pm$ 0.04, respectively.

When we compare $M_{K_{s}}^{\rm Bump}$ with the CG97 metallicity scale
[Fe/H]$_{\rm CG97}$ or the $\alpha$-element enhanced global
metallicity scale [M/H] given in Table 2, we find clear
quadratic correlations as shown in Figure 4. The relations expressed in
equations (5a) and (5b) imply that as metallicity decreases, the
luminosity of the RGB bump becomes brighter, which is an identical
result to  that of Ferraro et al. (2000).

\placefigure{fig4}

\begin{mathletters}
\begin{eqnarray}
M_{K_{s}}^{\rm Bump} = (0.26 \pm 0.16){\rm [Fe/H]}_{\rm CG97}^{2} +
(1.43 \pm 0.45){\rm [Fe/H]}_{\rm CG97} - (0.35 \pm 0.11) \\
M_{K_{s}}^{\rm Bump} = (0.33 \pm 0.18){\rm [M/H]}^{2} + (1.51 \pm
0.44){\rm [M/H]} - (0.51 \pm 0.11)
\end{eqnarray}
\end{mathletters}

The photometric data of Ferraro et al. (2000) were homogeneous, making use of
Glass' standard stars (Ferraro et al. 1994; Montegriffo et al.
1995). The 2MASS photometric data are also homogeneous and have
undergone Global Photometric Calibration (Nikolaev et al. 2000).
Therefore, if we estimate the photometric zero point differences between them
we can combine the two sets of photometric data
by correcting for the zero point difference. Comparison of RGB bump
positions of clusters common to both works can be used for estimation
of the zero point difference. However, since the two works were conducted
in different photometric systems, we first have to render the photometric
data in the same sustem, that is to say in the 2MASS system.

We transformed Ferraro et al.'s (2000) results into the 2MASS system
using equation (A1) of Carpenter (2001). Equation (A1) of
Carpenter (2001) can transform the Bessell \& Brett (1988) system
into the 2MASS system. The Bessell \& Brett system is the homogenized Glass
system, and the photometric data of Ferraro et al. (2000) are in the Glass
system since the data are standardized by Glass' standard stars. So,
we transformed the data of Ferraro et al. (2000) in the Glass system into the
2MASS system using equation (A1) of Carpenter (2001).

The 3 GGCs of this work overlap with those of Ferraro et al.
(2000), and the RGB bump luminosities of the two works are listed
in Table 3. We list the luminosities of the RGB bumps in the Glass
system which are original and those in the 2MASS system after
transformation from the Glass system for the 8 GGCs of
Ferraro et al. (2000) in Table 4.
According to Table 3, in the cases of M15 (NGC 7078) and 47 Tuc,
the RGB bump positions in this work are fainter by 0.09 mag than those of
Ferraro et al. (2000), and in the case of M107,
the RGB bump position in this work is brighter by 0.01 mag
than that of Ferraro et al. (2000).
However, since these differences are smaller than the errors of
the data, we can neglect the zero point difference between the two works.
Therefore, we combined the data in the two works without applying
any zero point correction. The results are shown in Figure 5.

\placetable{tbl-3}

\placetable{tbl-4}

\placefigure{fig5}

The sample of 8 GGCs in Ferraro et al. (2000) covers a wide range in
metallicity despite its small size. While our sample is a little larger,
comprising 11 GGCs, it covers a limited range in metallicity.
Therefore, the combined data set increases the sample of GGCs
to 16 and covers a wider range in metallicity than that of the
sample GGCs of our work. We have found  clear relations
between GGC metallicity and luminosity of the RGB bump, as shown in
Figure 5. By taking error-weighted mean averages for the 3 common GGCs
and combining the remaining data we have the reduced
regression equations (6a) and (6b) and we  plot the results
in Figure 5 as solid lines.

\begin{mathletters}
\begin{eqnarray}
M_{K_{s}}^{\rm Bump} = (0.05 \pm 0.06){\rm [Fe/H]}_{\rm CG97}^{2}
+ (0.89 \pm 0.12){\rm [Fe/H]}_{\rm CG97} - (0.67 \pm 0.10) \\
M_{K_{s}}^{\rm Bump} = (0.07 \pm 0.06){\rm [M/H]}^{2} + (0.93 \pm
0.11){\rm [M/H]} - (0.78 \pm 0.10)
\end{eqnarray}
\end{mathletters}

Dashed lines are from equations given in Figure 13 of
Ferraro et al. (2000) transformed into the 2MASS system and dot-dashed
lines are from equations (5a) and (5b). Although dashed lines and
dot-dashed lines are not exactly coincident with each other, their forms
are very similar. The first terms of equations (6a) and (6b) are negligible
if the errors in the parameters are considered. Therefore solid lines are
nearly linear and
deviate from the dashed lines slightly at the metal rich ends. This is due to
metallicity revision of NGC 6553 and NGC 6528. Their original values 
in Ferraro et al. (2000), which were extrapolated to be
[Fe/H]$_{\rm CG97}$ $= -0.44$ and $-0.38$, and [M/H] $= -0.36$
and $-0.31$, respectively, have been revised to [Fe/H]$_{\rm CG97}$
$= -0.06$ and 0.07, and [M/H] $=$ 0.08 and 0.22, respectively.

The overall trend for the RGB bump positions of GGCs to become
brighter with decreasing metallicity has been supported by
 many theoretical models derived from the first
such model established by  Sweigart (1978).
However, there is also a moderate
helium abundance dependency (Sweigart 1978) and a weak age dependency
(Ferraro et al 1999; Yi et al. 2001) of RGB bump positions for a given
metallicity.

According to Yi et al. (2001) and Ferraro et al. (1999), the RGB bump luminosity
varies with metallicity by $\Delta{M_{V}} \over \Delta{\rm [Fe/H]}$
$\approx$ 0.96 in the metallicity range [Fe/H] $=$ $-2.3$
$\sim$ 0.0 and in the age range of 7 Gyr $\sim$ 16 Gyr.
Yun \& Lee (1979) found that RGB bump bolometric luminosity varied with
[Fe/H] according to a nearly constant relation in the helium abundance range $Y$ $=$ 0.1 $\sim$ 0.3.
in the theoretical luminosity function analysis
of 46 RGB models of Sweigart \& Gross (1978).
Therefore, we assume that the RGB bump luminosity varies with [Fe/H] constantly
by $\Delta{M_{V}} \over \Delta{\rm [Fe/H]}$
$\approx$ 0.96 in the metallicity range [Fe/H] $=$ $-2.3$ $\sim$ 0.0, in
the age range of 7 Gyr $\sim$ 16 Gyr, and in the helium abundance range
$Y$ $=$ 0.1 $\sim$ 0.3. The total luminosity change of RGB bump has 
been found to be
$\sim$1.83 in $M_{V}$ (Ferraro et al. 1999) and $\sim$1.76 in $M_{K_{s}}$
(Ferraro et al. 2000) in the same metallicity range.
So, we derive the RGB bump luminosity variation in $M_{K_{s}}$ with [Fe/H]
as $\Delta{M_{K_{s}}} \over \Delta{\rm [Fe/H]}$ $\approx$ $1.76 \over 1.83$$
\Delta{M_{V}} \over \Delta{\rm [Fe/H]}$ $\approx$ 0.92
in the metallicity range [Fe/H] $= -2.3 \sim$ 0.0, in the age range of 7 Gyr
$\sim$ 16 Gyr, and in the helium abundance range $Y$ $= 0.1 \sim$ 0.3.

According to the theoretical luminosity function analysis of 46 RGB models
of Sweigart \& Gross (1978) by Yun \& Lee (1979), when mass
${\cal M} = 0.9$ ${\cal M}_{\odot}$, the
RGB bump bolometric luminosity variation with helium abundance $Y$
is $\Delta{m} \over \Delta{Y}$ $\approx -5.0$ in the
helium abundance range $Y$ $= 0.1 \sim 0.3$ and in the metallicity
range [Fe/H] $= -2.4 \sim -0.3$. Here $m$ is bolometric
magnitude which is proportional to $-2.5\times{\rm log}{(L/L_{\odot})}$ and
taken to be 0 at the RGB tip. Comparison of the theoretical RGB bump
bolometric variation rate by Yun \& Lee (1979) with the theoretical RGB
bump luminosity variation rate in $M_{V}$ by Yi et al. (2001) gives
$\Delta{m} \approx 1.21\Delta{M_{V}} \approx 1.26\Delta{M_{K_{s}}}$. So,
when mass ${\cal M} = 0.9$ ${\cal M}_{\odot}$, the RGB bump luminosity variation
in $M_{K_{s}}$ with
helium abundance $Y$ is found to be $\Delta{M_{K_{s}}} \over \Delta{Y}$
$\approx$ $-4.0$ in the helium abundance range $Y$ $= 0.1 \sim 0.3$ and
in the metallicity range [Fe/H] $= -2.4 \sim -0.3$.

According to Yi et al. (2001) and Ferraro et al. (1999), the RGB bump
luminosity variation in $M_{V}$ with age is $\Delta{M_{V}}
\over \Delta(t_{9})$ $\approx$ 0.04 in the age range of 7 Gyr $\sim$
16 Gyr and in the metallicity range [Fe/H] $= -2.3 \sim$ 0.0, where
$t_{9}$ represents age in Gyr unit. Since the relation between the total
luminosity changes of the RGB bump in $M_{V}$ and in $M_{K_{s}}$ is
$\Delta{M_{V}} \approx 1.04\Delta{M_{K_{s}}}$, RGB bump luminosity
variation in $M_{K_{s}}$ with age is found to be $\Delta{M_{K_{s}}} \over
\Delta{(t_{9})}$ $\approx$ 0.04 in the age range of 7 Gyr $\sim$ 16 Gyr
and in the metallicity range [Fe/H] $= -2.3 \sim$ 0.0.

Applying equation (11) of Buzzoni et al. (1983) for the relation between $R$
($N_{\rm HB}/N_{\rm RGB}$, the ratio of the number of HB stars to the
number of RGB stars brighter than the HB) and helium abundance $Y$ to $R$
values of 26 GGCs (Zoccali et al. 2000), the maximum range of helium
abundance in GGCs at a given [Fe/H] is $\sim$0.06 ($\Delta{Y}$ $=
\pm0.03$). However, according to the $Y$ data of Sandquist (2000) for
43 GGCs, the maximum range of helium abundance in GGCs at a given [Fe/H] is
$\sim$0.10 ($\Delta{Y}$ $= \pm0.05$).
So, the maximum RGB bump luminosity variation in $M_{K_{s}}$ caused by
helium abundance variation in GGCs at a given [Fe/H] is found to be
$\Delta{M_{K_{s}}} \approx -4.0\Delta{Y} \approx \pm$0.20 mag.

 However, the maximum age spread of GGCs at a given [Fe/H] is
$\sim$6 Gyr ($\Delta{(t_{9})} = \pm$3 Gyr) according to Chaboyer et al. (1996), Richer et al. (1996), Salaris \&
Weiss (1997), Buonanno et al. (1998), Rosenberg et al. (1999), and
 VandenBerg (2000). So, the maximum RGB bump luminosity
variation in $M_{K_{s}}$ caused by age variation in GGCs at a given
[Fe/H] is found to be $\Delta{M_{K_{s}}} \approx 0.04\Delta{(t_{9})}$
$\approx \pm$0.12 mag.

Therefore, the maximum variation of RGB bump luminosity in $M_{K_{s}}$ with
helium abundance and age spreads in GGCs at a given [Fe/H] is
$\Delta{M_{K_{s}}} \approx \pm$$\sqrt{0.20^2 + 0.12^2}$ $\approx$
$\pm$0.24 mag. This implies that the relations between the
absolute magnitude of the RGB bump ($M_{K_{s}}^{\rm Bump}$) and metallicity
${\rm [Fe/H]}_{\rm CG97}$ or [M/H] derived in this work can be used to
determine distance moduli of other GGCs with an uncertainty of approximately
$\Delta{M_{K_{s}}}$ $\approx \pm$0.24 mag at a given metallicity.

\section{Summary}

Using photometric data from the 2MASS second incremental release point
source catalog we found RGB bump features in 13 GGCs in $K_{s}$
vs. ($J-K_{s}$) CMDs and measured accurate positions of RGB bumps
for 11 GGCs excluding M69 and $\omega$ Cen. We have found
clear relations between $M_{K_{s}}^{\rm Bump}$
and metallicity [Fe/H]$_{\rm CG97}$ or [M/H], thereby
independently confirming the results of Ferraro et al. (2000).

Combining the present sample and that of Ferraro et al. (2000),
we extend the number of GGCs to 16 whose RGB
bump positions have been measured, and determined more robust
correlations between the absolute
magnitudes of RGB bumps ($M_{K_{s}}^{\rm Bump}$) and metallicity
[Fe/H]$_{\rm CG97}$ or [M/H]. Furthermore, these equations can be used to
determine distance moduli of other GGCs with an uncertainty of
$\Delta{M_{K_{s}}}$ $\approx \pm$0.24 mag at a given metallicity.

\acknowledgments
This publication makes use of data products from the Two Micron All 
Sky Survey, which is a joint project of the University of Massachusetts
and the Infrared Processing and Analysis Center/California Institute of 
Technology, funded by the National Aeronautics and Space Administration
and the National Science Foundation.
This work is supported by the BK21 project of Korea through School of
Earth and Environmental Sciences (Astronomy Program), Seoul National
University.

\clearpage

\begin{figure}
\figurenum{1} \epsscale{1} \plotone{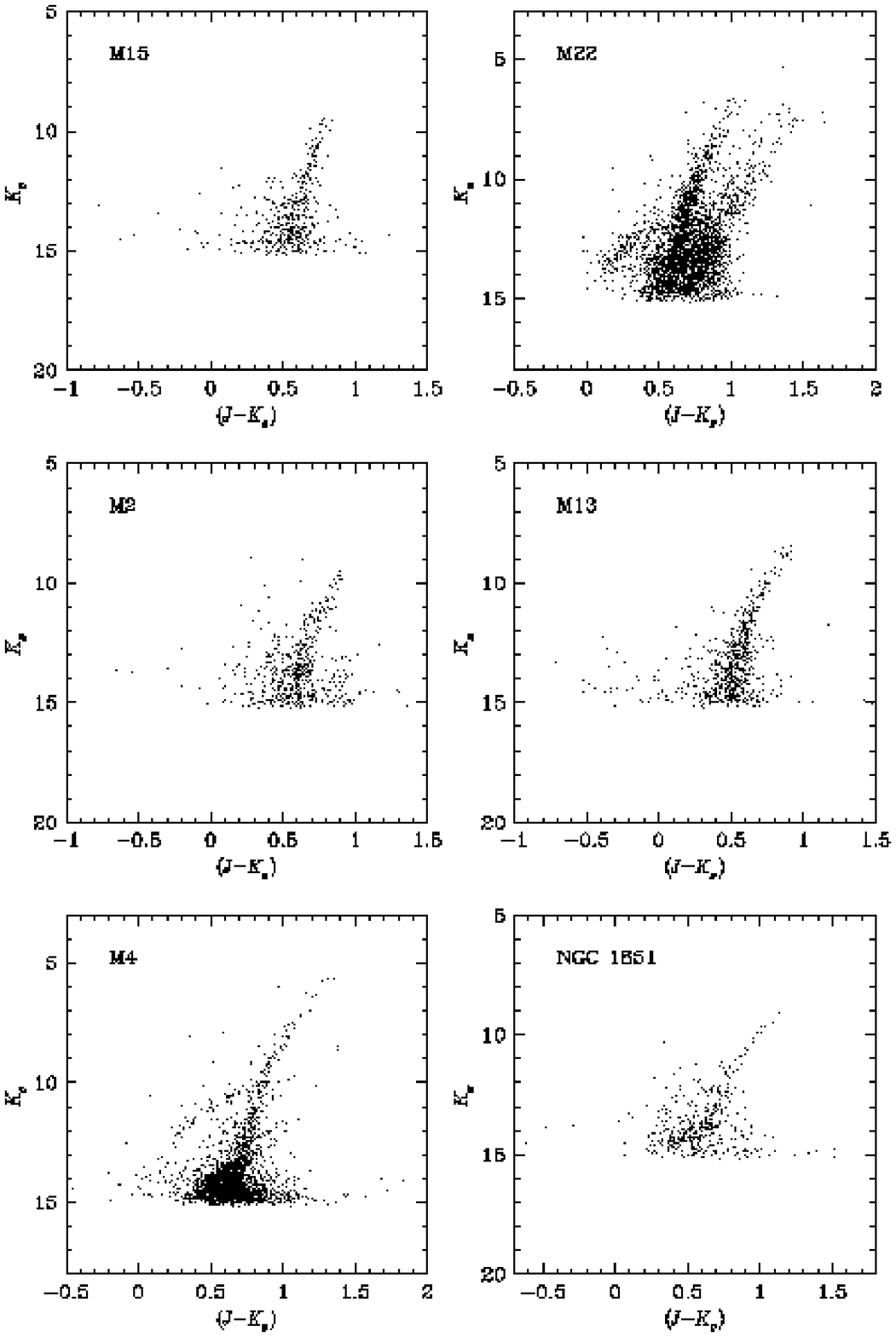}
\caption{$K_{s}$ vs.
($J-K_{s}$) CMDs of GGCs whose RGB bump positions are
accurately measured.}
\label{fig1}
\end{figure}

\begin{figure}
\figurenum{1} \epsscale{1} \plotone{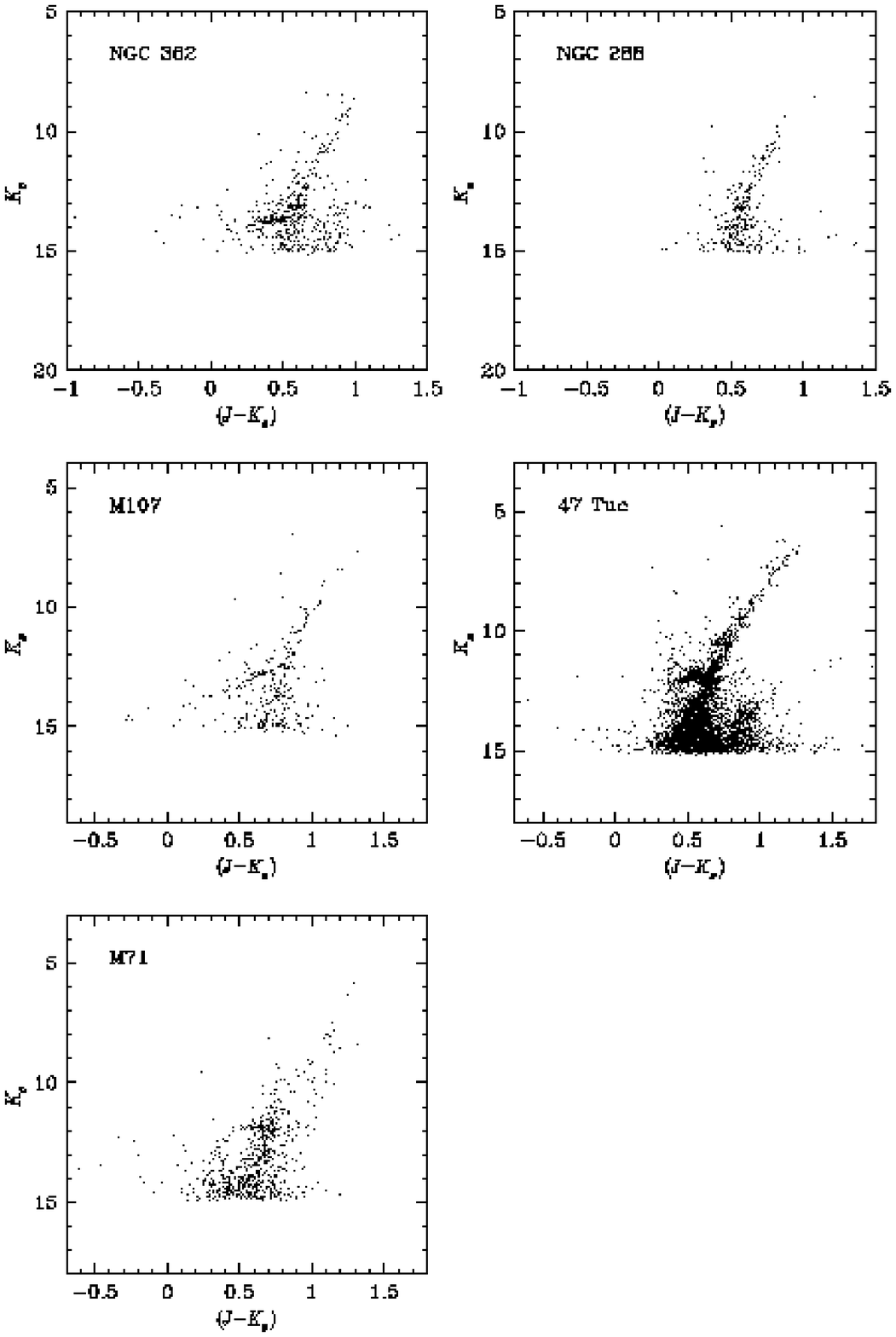}
\caption{{\it Continued}.}
\label{fig1}
\end{figure}

\begin{figure}
\figurenum{2} \epsscale{1} \plotone{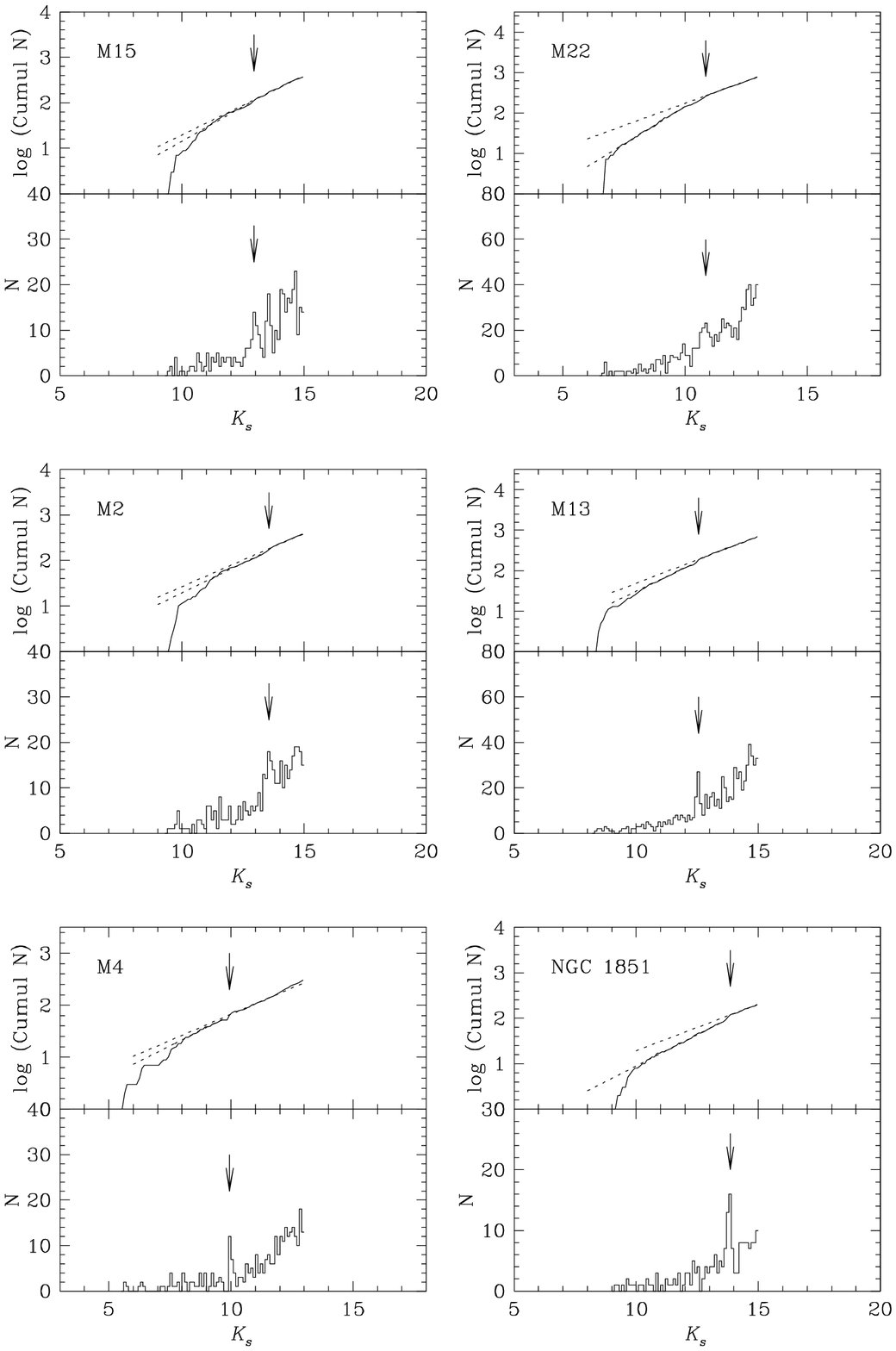}
\caption{Luminosity
functions of GGCs in Fig. 1. In each diagram the upper part is the
integral luminosity function and the lower part is the differential
luminosity function. Arrow in each diagram indicates RGB bump
position. The dashed lines in the upper part of each diagram are
linear fits to the regions above and below the RGB bump.}
\label{fig2}
\end{figure}

\begin{figure}
\figurenum{2} \epsscale{1} \plotone{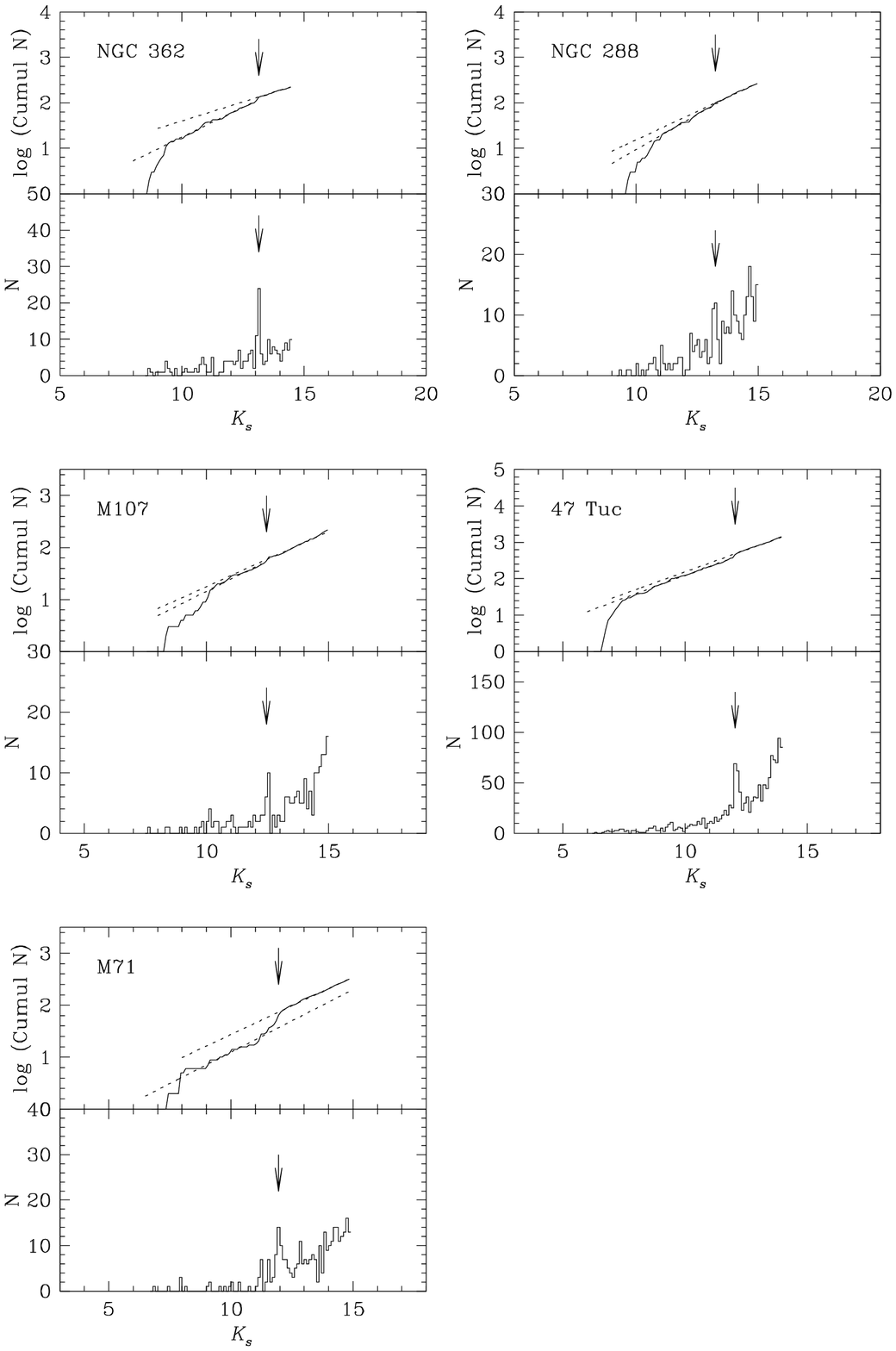}
\caption{{\it Continued}.}
\label{fig2}
\end{figure}

\begin{figure}
\figurenum{3} \epsscale{1} \plotone{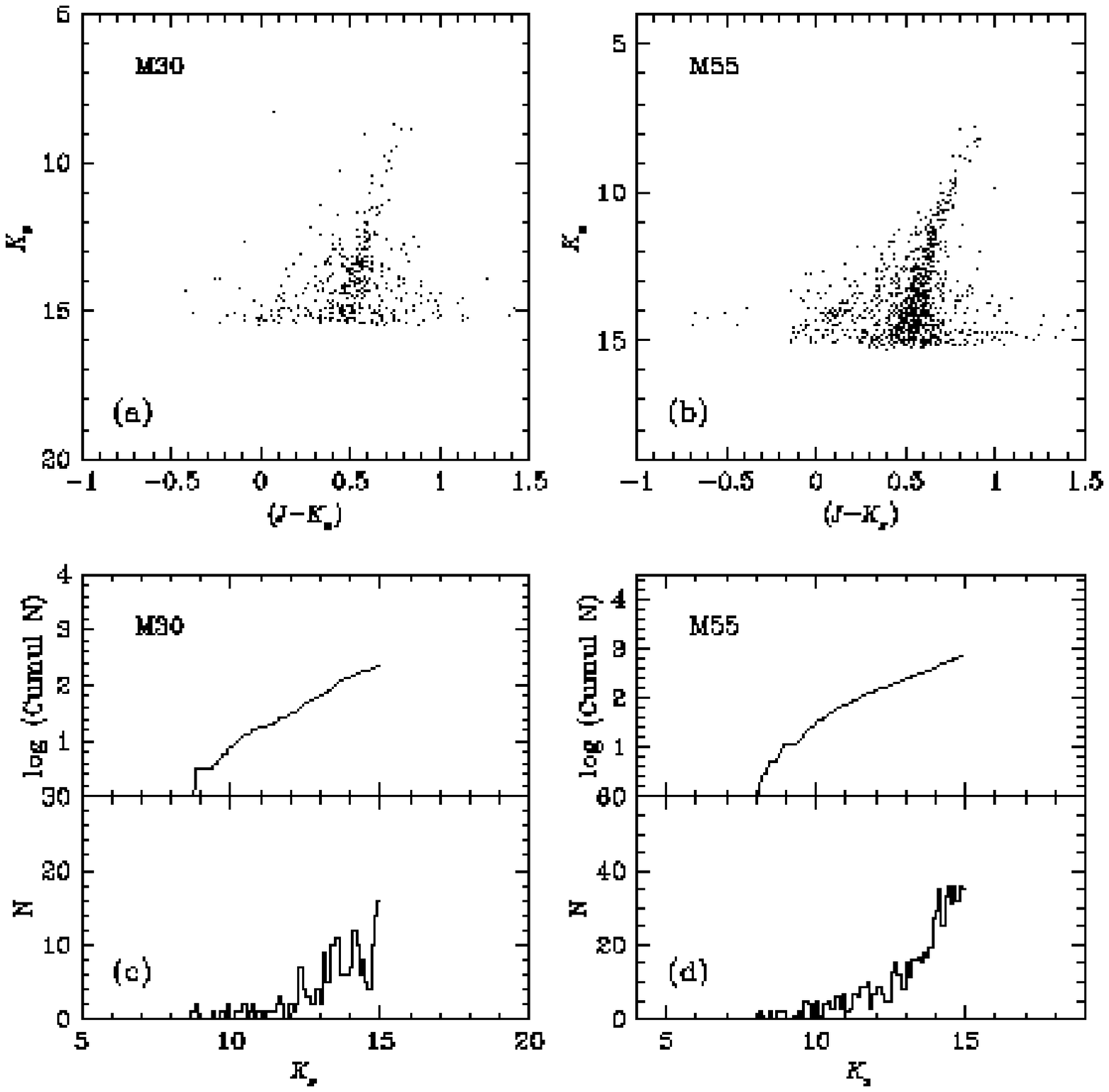}
\caption{({\it a})
$K_{s}$ vs. $(J-K_{s})$ CMD of M30. ({\it b}) $K_{s}$ vs. $(J-K_{s})$
CMD of M55. ({\it c}) The upper part is the integral luminosity function
and the lower part is the differential luminosity function of M30.
({\it d}) Same as Fig. 3{\it c} but for M55.}
\label{fig3}
\end{figure}

\begin{figure}
\figurenum{4} \epsscale{0.5} \plotone{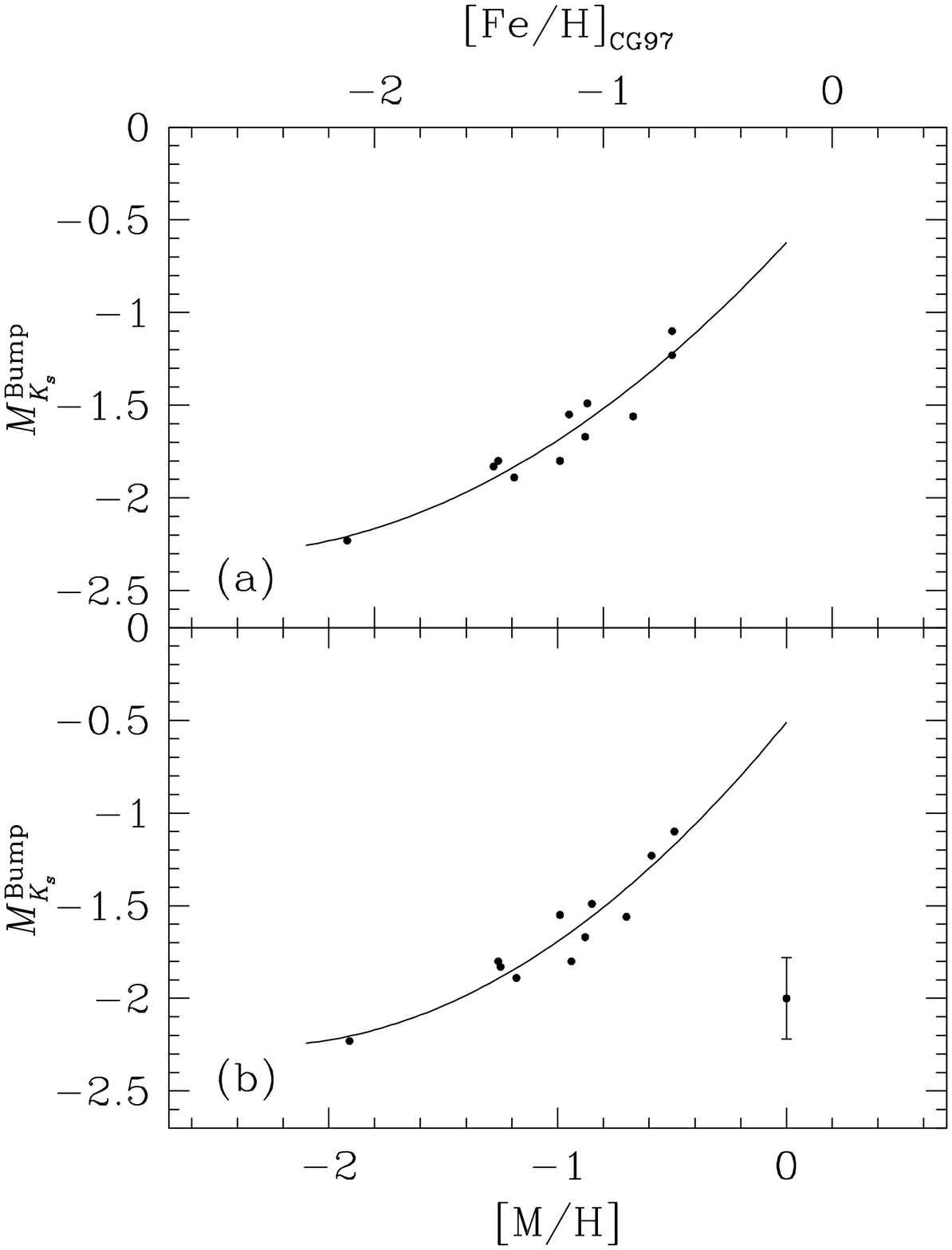}
\caption{({\it a})
Relation between $M_{K_{s}}^{\rm Bump}$ and [Fe/H]$_{\rm CG97}$
derived in this work. ({\it b}) Relation between $M_{K_{s}}^{\rm
Bump}$ and [M/H] derived in this work. Solid line in each diagram
is from eqs. (5a) and (5b). Small filled circle with error bar in
the lower right corner of Fig. 4{\it b} shows typical error sizes
of the values of small filled circles in Fig. 4.}
\label{fig4}
\end{figure}

\begin{figure}
\figurenum{5} \epsscale{0.5} \plotone{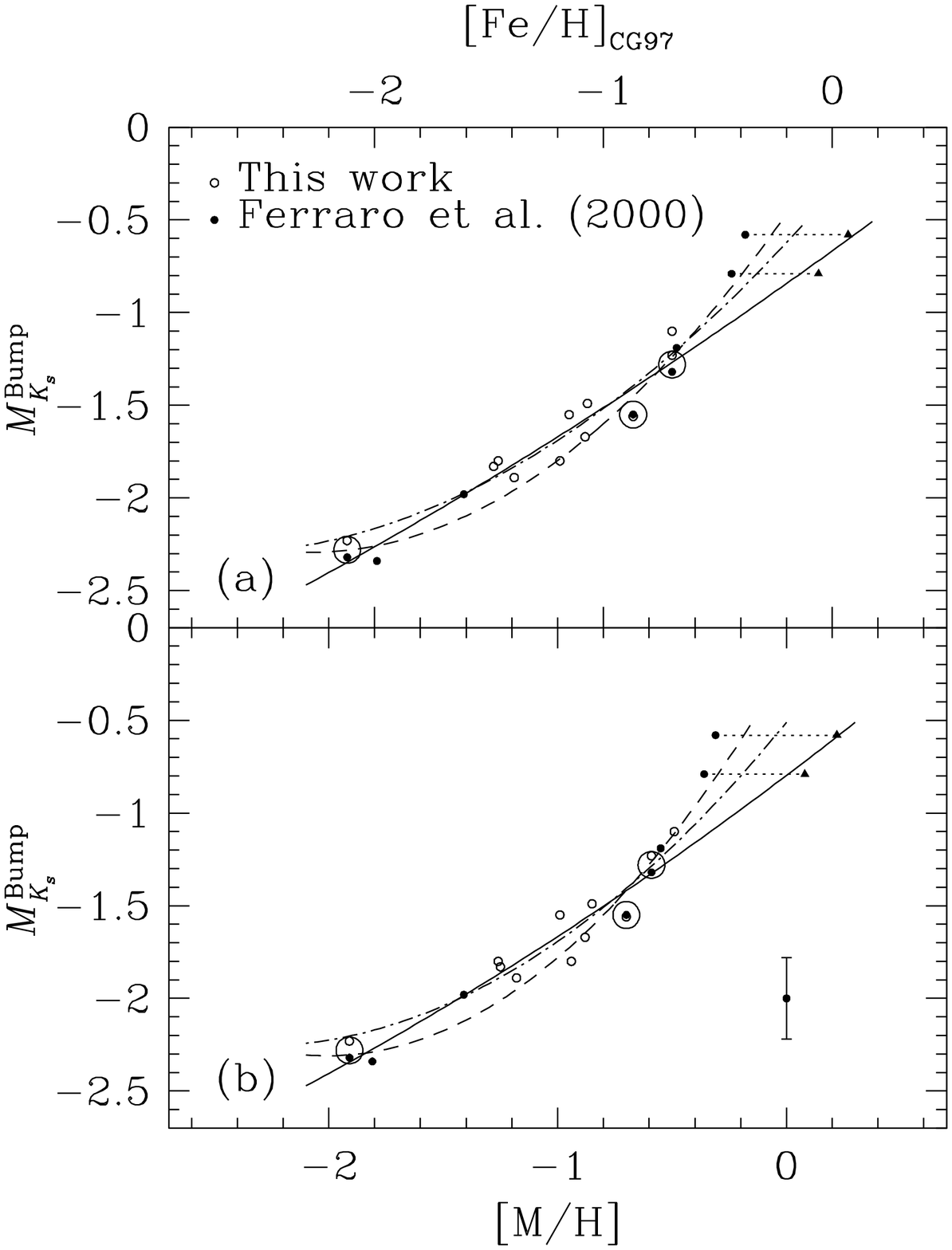} \caption{({\it a})
Relation between $M_{K_{s}}^{\rm Bump}$ and [Fe/H]$_{\rm CG97}$
combining results of this work and those of Ferraro et al. (2000)
transformed into 2MASS system. ({\it b}) Relation between
$M_{K_{s}}^{\rm Bump}$ and [M/H] combining results of this work
and those of Ferraro et al. (2000) transformed into 2MASS system.
Small open circles represent values from this work and small
filled circles represent those of Ferraro et al. (2000)
transformed into 2MASS system. Small filled triangles represent
new values for NGC 6553 and NGC 6528 caused by metallicity revision
due to high resolution spectroscopic measurements by Carretta et al.
(2001) and are connected to the original values of Ferraro et al.
(2000) by dotted lines. Two small circles in one larger
circle represent different values of same GGC common in both
works and larger circle  represents average of values of
two small circles enclosed in the larger circle. In the case of
M107 the two values differ only by 0.01 mag so they are nearly
indistinguishable. Solid line in each diagram is from eqs.
(6a) and (6b) and dashed line in each diagram is from equations
given in each panel of Fig. 13 of Ferraro et al. (2000)
transformed into 2MASS system and dot-dashed line in each diagram
is from eqs. (5a) and (5b) as given in Fig. 4{\it a} and 4{\it b}.
Small filled circle with error bar
in the lower right corner of Fig. 5{\it b} shows typical error
sizes of the values of small open and filled circles and small filled
triangles in Fig. 5.}
\label{fig5}
\end{figure}

\clearpage

\begin{deluxetable}{lrcrlrc}
\tablenum{1} \tablecolumns{7} \tablecaption{Parameters of M22 and
M2. \label{tbl-1}} \tablewidth{0pt} \tablehead{
\multicolumn{3}{c}{M22} & \colhead{} &
\multicolumn{3}{c}{M2} \\
\cline{1-3} \cline{5-7} \\
\colhead{Parameter} & \colhead{Value} & \colhead{Reference} &
\colhead{} & \colhead{Parameter} & \colhead{Value} &
\colhead{Reference}}
\startdata
${\rm [Fe/H]}_{\rm CG97}$ & $-$1.48  & 1 & &  ${\rm [Fe/H]}_{\rm CG97}$
& $-$1.46 & 5 \\
${\rm [M/H]}$             & $-$1.25  & 2 & &  ${\rm [M/H]}$
& $-$1.26 & 2 \\
${<}V_{\rm HB}{>}$        &   14.10  & 3 & &  ${<}V_{\rm HB}{>}$
& 15.93 & 5 \\
$V_{\rm ZAHB}$            &   14.16  & 2 & &  $V_{\rm ZAHB}$
& 15.99 & 2 \\
$M_{V}^{\rm ZAHB}$        &    0.59  & 2 & &  $M_{V}^{\rm ZAHB}$
&  0.59 & 2 \\
$E(B-V)$                  &    0.34  & 4 & &  $E(B-V)$
&  0.02 & 5 \\
$(m-M)_{0}^{\rm CG97}$    &   12.52  & 2 & &  $(m-M)_{0}^{\rm
CG97}$
& 15.34 & 2 \\
\enddata

\tablerefs{ (1) Carretta \& Gratton (1997); (2) this work;
(3) Peterson \& Cudworth (1994); (4) Harris (1996);
(5) Lee \& Carney (1999)}

\end{deluxetable}

\clearpage

\begin{deluxetable}{lcccccc} \rotate
\tablenum{2} \tablecaption{Parameters and derived RGB Bump
Positions of Program GGCs. \label{tbl-2}} \tablewidth{0pt}
\tablehead{ \colhead{Cluster} & \colhead{[Fe/H]$_{\rm CG97}$} &
\colhead{[M/H]} & \colhead{$E(B-V)$} & \colhead{$(m-M)_{0}$} &
\colhead{$K_{s}^{\rm Bump}$} & \colhead{$M_{K_{s}}^{\rm Bump}$} }
\startdata M15 (NGC 7078)   &$-$2.12 &$-$1.91 &0.09 &15.15 &12.95
$\pm$ 0.05 &$-$2.23
$\pm$ 0.21 \\
M22 (NGC 6656)   &$-$1.48 &$-$1.25 &0.34 &12.52 &10.80 $\pm$ 0.10
&$-$1.83
$\pm$ 0.22 \\
M2 (NGC 7089)    &$-$1.46 &$-$1.26 &0.02 &15.34 &13.55 $\pm$ 0.05
&$-$1.80
$\pm$ 0.21 \\
M13 (NGC 6205)   &$-$1.39 &$-$1.18 &0.02 &14.43 &12.55 $\pm$ 0.05
&$-$1.89
$\pm$ 0.21 \\
M4 (NGC 6121)    &$-$1.19 &$-$0.94 &0.36 &11.68 &10.00 $\pm$ 0.10
&$-$1.80
$\pm$ 0.22 \\
NGC 362          &$-$1.15 &$-$0.99 &0.05 &14.68 &13.15 $\pm$ 0.05
&$-$1.55
$\pm$ 0.21 \\
NGC 1851         &$-$1.08 &$-$0.88 &0.02 &15.46 &13.80 $\pm$ 0.10
&$-$1.67
$\pm$ 0.22 \\
NGC 288          &$-$1.07 &$-$0.85 &0.03 &14.73 &13.25 $\pm$ 0.05
&$-$1.49
$\pm$ 0.21 \\
M107 (NGC 6171)  &$-$0.87 &$-$0.70 &0.33 &13.95 &12.50 $\pm$ 0.10
&$-$1.56
$\pm$ 0.22 \\
47 Tuc (NGC 104) &$-$0.70 &$-$0.59 &0.04 &13.32 &12.10 $\pm$ 0.10
&$-$1.23
$\pm$ 0.22 \\
M71 (NGC 6838)   &$-$0.70 &$-$0.49 &0.25 &12.97 &11.95 $\pm$ 0.05
&$-$1.10
$\pm$ 0.21 \\
\enddata

\tablecomments{Metallicities, interstellar reddenings, and
distance moduli of GGCs are taken from Table 1 of Ferraro et al.
(1999) except for M22 and M2 whose values are taken from
Table 1 of this work. More details are described in the text.}

\end{deluxetable}

\clearpage

\begin{deluxetable}{lcccccccc}
\tablenum{3} \tablecolumns{9} \tablecaption{Comparison of RGB
Bump Positions of 3 overlapping GGCs in Ferraro et al. (2000) and
in This Work. \label{tbl-3}} \tablewidth{0pt} \tablehead{
\colhead{} & \colhead{} & \multicolumn{2}{c}{Ferraro et al.
(2000)} & \colhead{} & \multicolumn{2}{c}{This work} & \colhead{}
&
\colhead{Average\tablenotemark{a}} \\
\cline{3-4} \cline{6-7} \cline{9-9} \\
\colhead{Cluster} & \colhead{} & \colhead{$K_{s}^{\rm Bump}$} &
\colhead{$M_{K_{s}}^{\rm Bump}$}   & \colhead{}               &
\colhead{$K_{s}^{\rm Bump}$}       & \colhead{$M_{K_{s}}^{\rm
Bump}$} & \colhead{} & \colhead{${<}M_{K_{s}}^{\rm Bump}{>}$}}
\startdata M15    &  &12.86 $\pm$ 0.05  &$-$2.32 $\pm$ 0.21 &
&12.95 $\pm$ 0.05 &
$-$2.23 $\pm$ 0.21 &   &$-$2.28 $\pm$ 0.15 \\
M107   &  &12.51 $\pm$ 0.05  &$-$1.55 $\pm$ 0.21 &  &12.50 $\pm$
0.10 &
$-$1.56 $\pm$ 0.22 &   &$-$1.55 $\pm$ 0.15 \\
47 Tuc &  &12.01 $\pm$ 0.05  &$-$1.32 $\pm$ 0.21 &  &12.10 $\pm$
0.10 &
$-$1.23 $\pm$ 0.22 &   &$-$1.28 $\pm$ 0.15 \\
\enddata

\tablecomments{Ferraro et al.'s (2000) original results in Glass
system were transformed into 2MASS system according to equation
(A1) of Carpenter (2001).} \tablenotetext{a}{Weighted mean
average by errors.}

\end{deluxetable}

\clearpage

\begin{deluxetable}{lcccc}
\tablenum{4} \tablecaption{Absolute Magnitudes of RGB Bumps of 8
GGCs derived by Ferraro et al. (2000). \label{tbl-4}}
\tablewidth{0pt} \tablehead{ \colhead{Cluster} &
\colhead{[Fe/H]$_{\rm CG97}$} & \colhead{[M/H]} &
\colhead{$M_{K}^{\rm Bump}$} & \colhead{$M_{K_{s}}^{\rm Bump}$} }
\startdata
M15 (NGC 7078)   &$-$2.12 &$-$1.91 &$-$2.28 $\pm$ 0.21 &$-$2.32 $\pm$ 0.21 \\
M68 (NGC 4590)   &$-$1.99 &$-$1.81 &$-$2.30 $\pm$ 0.21 &$-$2.34 $\pm$ 0.21 \\
M55 (NGC 6809)   &$-$1.61 &$-$1.41 &$-$1.94 $\pm$ 0.21 &$-$1.98 $\pm$ 0.21 \\
M107 (NGC 6171)  &$-$0.87 &$-$0.70 &$-$1.51 $\pm$ 0.21 &$-$1.55 $\pm$ 0.21 \\
47 Tuc (NGC 104) &$-$0.70 &$-$0.59 &$-$1.28 $\pm$ 0.21 &$-$1.32 $\pm$ 0.21 \\
M69 (NGC 6637)   &$-$0.68 &$-$0.55 &$-$1.15 $\pm$ 0.21 &$-$1.19 $\pm$ 0.21 \\
NGC 6553         &$-$0.06\tablenotemark{a} &$+$0.08\tablenotemark{b} &
$-$0.75 $\pm$ 0.22 &$-$0.79 $\pm$ 0.22 \\
NGC 6528         &$+$0.07\tablenotemark{a} &$+$0.22\tablenotemark{b} &
$-$0.54 $\pm$ 0.22 &$-$0.58 $\pm$ 0.22 \\
\enddata

\tablecomments{Cols. (2) and (3) list metallicities taken from
Table 1 of Ferraro et al. (2000) except for NGC 6553 and NGC 6528.
Col. (4) lists absolute
magnitudes of RGB bumps in Glass system and col. (5) lists those
values transformed into 2MASS system according to eq. (A1) of
Carpenter (2001).} \tablenotetext{a}{These values were taken from
Carretta et al. (2001).} \tablenotetext{b}{These values were calculated
by equation (3) in $\S$ 3.1 and more details are described in the text.}

\end{deluxetable}


\begin{thebibliography}{}
\bibitem[Armandroff and Zinn(1988)]{arm88} Armandroff, T. E., and Zinn, R.
    1988, \aj, 96, 92
\bibitem[Bessell and Brett(1988)]{bes88} Bessell, M. S., and Brett, J. M.
    1988, \pasp, 100, 1134
\bibitem[Bessell et al.(1998)]{bes98} Bessell, M. S., Castelli, F., and
    Plez, B. 1998, \aap, 333, 231
\bibitem[Brocato et al.(1996)]{bro96} Brocato, E., Buonanno, R., Malakhova,
    Y., and Piersimoni, A. M. 1996, \aap, 311, 778
\bibitem[Buonanno et al.(1998)]{buo98} Buonanno, R., Corsi, C. E., Pulone,
    L., Fusi Pecci, F., and Bellazzini, M. 1998, \aap, 333, 505
\bibitem[Buzzoni et al.(1983)]{buz83} Buzzoni, A., Fusi Pecci, F.,
    Buonanno, R., and Corsi, C. E. 1983, \aap, 128, 94
\bibitem[Carney(1996)]{car96} Carney, B. W. 1996, \pasp, 108, 900
\bibitem[Carpenter(2001)]{carp01} Carpenter, J. M. 2001, \aj, 121, 2851
\bibitem[Carretta et al.(2001)]{car01} Carretta, E., Cohen, J. G.,
    Gratton, R. G., and Behr, B. B. 2001, \aj, 122, 1469
\bibitem[Carretta and Gratton(1997)]{car97} Carretta, E., and Gratton, R. G.
    1997, \aaps, 121, 95
\bibitem[Chaboyer et al.(1996)]{cha96} Chaboyer, B., Demarque, P., and
    Sarajedini, A. 1996, \apj, 459, 558
\bibitem[Cohen et al.(1999)]{coh99} Cohen, J. G., Gratton, R. G., Behr,
    B. B., and Carretta, E. 1999, \apj, 523, 739
\bibitem[Cudworth(1985)]{cud85} Cudworth, K. M. 1985, \aj, 90, 65
\bibitem[Ferraro et al.(1994)]{fer94} Ferraro, F. R., Fusi Pecci, F.,
    Guarnieri, M. D., Moneti, A., Origlia, L., and Testa, V. 1994,
    \mnras, 266, 829
\bibitem[Ferraro et al.(1999)]{fer99} Ferraro, F. R., Messineo, M., Fusi
    Pecci, F., De Palo, M. A., Straniero, O., Chieffi, A., and Limongi, M.
    1999, \aj, 118, 1738
\bibitem[Ferraro et al.(2000)]{fer00} Ferraro, F. R., Montegriffo, P.,
    Origlia, L., and Fusi Pecci, F. 2000, \aj, 119, 1282
\bibitem[Fusi Pecci et al.(1990)]{fus90} Fusi Pecci, F., Ferraro, F. R.,
    Crocker, D. A., Rood, R. T., and Buonanno, R. 1990, \aap, 238, 95
\bibitem[Girardi et al.(2000)]{gir00} Girardi, L., Bressan, A., Bertelli, G.,
    and Chiosi, C. 2000, \aaps, 141, 371
\bibitem[Harris(1996)]{har96} Harris, W. E. 1996, \aj, 112, 1487
\bibitem[He et al.(1995)]{hel95} He, L., Whittet, D. C. B., Kilkenney, D.,
    and Spencer Jones, J. H. 1995, \apjs, 101, 336
\bibitem[Iben(1968)]{ibe68} Iben, I. 1968, \nat, 220, 143
\bibitem[King et al.(1985)]{kin85} King, C. R., Da Costa, G. S., and
    Demarque, P. 1985, \apj, 299, 674
\bibitem[Lee and Carney(1999)]{lee99} Lee, J.-W., and Carney, B. W. 1999, \aj,
    118, 1373
\bibitem[Lee et al.(1999)]{leey99} Lee, Y.-W., Joo, J.-M., Sohn, Y.-J., Rey,
    S.-C., Lee, H.-C., and Walker, A. R. 1999, \nat, 402, 55
\bibitem[Montegriffo et al.(1995)]{mon95} Montegriffo, P., Ferraro, F. R.,
    Fusi Pecci, F., and Origlia, L. 1995, \mnras, 276, 739
\bibitem[Nikolaev et al.(2000)]{nik00} Nikolaev, S., Weinberg, M. D.,
    Skrutskie, M. F., Cutri, R. M., Wheelock, S. L., Gizis, J. E., and
    Howard, E. M. 2000, \aj, 120, 3340
\bibitem[Ortolani et al.(1990)]{ort90} Ortolani, S., Barbuy, B., and Bica, E.
    1990, \aap, 236, 362
\bibitem[Pancino et al.(2000)]{pan00} Pancino, E., Ferraro, F. R.,
    Bellazzini, M., Piotto, G., and Zoccali, M. 2000, \apjl, 534, L83
\bibitem[Peterson and Cudworth(1994)]{pet94} Peterson, R. C., and Cudworth,
    K. M. 1994, \apj, 411, 103
\bibitem[Richer et al.(1996)]{ric96} Richer, H. B., et al. 1996, \apj, 463, 602
\bibitem[Rosenberg et al.(1999)]{ros99} Rosenberg, A., Saviane, I., Piotto,
    G., and Aparicio, A. 1999, \aj, 118, 2306
\bibitem[Salaris and Cassisi(1996)]{sal96} Salaris, M., and Cassisi, S. 1996,
    \aap, 305, 858
\bibitem[Salaris and Weiss(1997)]{sal97} Salaris, M., and Weiss, A. 1997,
    \aap, 327, 107
\bibitem[Sandquist(2000)]{san00} Sandquist, E. L. 2000, \mnras, 313, 571
\bibitem[Sarajedini and Norris(1994)]{sar94} Sarajedini, A., and Norris, J.
    E. 1994, \apjs, 93, 161
\bibitem[Saviane et al.(1998)]{sav98} Saviane, I., Piotto, G., Fagotto, F.,
    Zaggia, S., Capaccioli, M., and Aparicio, A. 1998, \aap, 333, 479
\bibitem[Sweigart(1978)]{swe78a} Sweigart, A. V. 1978, in IAU Symp. 80,
    The HR Diagram, ed. A. G. D. Philip \& D. S. Hayes (Dordrecht: Reidel),
    333
\bibitem[Sweigart and Gross(1978)]{swe78b} Sweigart, A. V., and Gross,
    P. G., 1978, \apjs, 36, 405
\bibitem[Thomas(1967)]{tho67} Thomas, H.-C. 1967, Z. Astrophys., 67, 420
\bibitem[VandenBerg(2000)]{van00} VandenBerg, D. A. 2000, \apjs, 129, 315
\bibitem[Yi et al.(2001)]{yis01} Yi, S., Demarque, P., Kim, Y.-C., Lee,
    Y.-W., Ree, C. H., Lejeune, T., and Barnes, S. 2001, \apjs, 136, 417
\bibitem[Yun and Lee(1979)]{yun79} Yun, H.-Y., and Lee, S.-W. 1979,
    J. Korean Astron. Soc., 12, 17
\bibitem[Zinn(1980)]{zin80} Zinn, R. 1980, \apjs, 42, 19
\bibitem[Zinn(1985)]{zin85} ---------. 1985, \apj, 293, 424
\bibitem[Zinn and West(1984)]{zin84} Zinn, R., and West, M. 1984, \apjs, 55, 45
\bibitem[Zoccali et al.(2000)]{zoc00} Zoccali, M., Cassisi, S., Bono, G.,
    Piotto, G., Rich, R. M., and Djorgovski, S. G. 2000, \apj, 538, 289
\end{thebibliography}
\end{document}